\documentstyle[seceq,preprint,epsf]{jpsj}
\newcommand{\state}{|\psi\rangle}

\newcommand{\slat}[1]{|\phi_#1\rangle}
\newcommand{\lstat}[1]{\langle\phi_#1|}
\newcommand{\rstat}[1]{|\phi_#1\rangle}
\def\runtitle{PIRG Studies 
on Magnetic and Metal-Insulator Transitions 
in the Two-Dimensional Hubbard Model}

\def\runauthor{Tsuyoshi {\sc Kashima} and Masatoshi {\sc Imada}}

\title{Path-Integral Renormalization Group Method for Numerical Study \\ on Ground States of Strongly Correlated Electronic Systems}

\author{ Tsuyoshi {\sc Kashima} and Masatoshi {\sc Imada}}
\inst{Institute for Solid State Physics,\\
University of Tokyo,5-1-5 Kashiwanoha,\\
Kashiwa, Chiba, 277-8581}

\recdate{\today}

\abst{A new efficient numerical algorithm for interacting fermion systems is proposed and examined  in detail. 
The ground state is expressed approximately by a linear combination of
numerically chosen basis states in a truncated Hilbert space. 
Two procedures lead to a better approximation. 
The first is a numerical renormalization, which optimizes the chosen
basis and projects onto the ground state within the fixed dimension,
$L$, of the Hilbert space.  
The second is an increase of the dimension of the truncated Hilbert
space, which enables the linear combination to converge to a better
approximation. 
The extrapolation $L\rightarrow\infty$ after the 
convergence removes the approximation error systematically. 
This algorithm does not suffer from the negative sign problem and can be
applied to systems in any spatial dimension and arbitrary lattice
structure. 
The efficiency is tested and the implementation explained 
for two-dimensional Hubbard models where Slater determinants are
employed as chosen basis. 
Our results with less than $400$ chosen basis indicate good accuracy 
within the errorbar of the best available results as those of the quantum 
Monte Carlo for energy and other physical quantities.  }

\kword{quantum simulation, strongly correlated electron systems, Hubbard model,numerical renormalization group}

\begin{document}
\sloppy
\maketitle
\section{Introduction}
In these decades, many numerical algorithms, such as exact
diagonalizations, quantum Monte Carlo and
density matrix renormalization group(DMRG), were proposed and were
applied to many strongly correlated electron systems. 
Though the exact diagonalization method is the most straightforward one, the
system size it can treat is smaller than that other methods can treat.
Quantum Monte Carlo(QMC) method is a powerful technique for correlated
electron systems and has been applied to various systems
\cite{QMC1,QMC2}. In some systems such as fermion
systems and frustrated spin systems, however, the QMC is known to suffer
from the negative sign problem. Namely, 
the cancellation of positive and negative Monte Carlo samples 
occurs and makes it practically impossible to estimate physical values 
in the presence of statistical and round-off errors.  
DMRG\cite{DMRG} is a very powerful 
numerical renormalization method which does not suffer from any sign
problem. However, because of the spatial renormalization process, DMRG
is known to be applied efficiently only to one-dimensional configurations.

Path-integral renormalization group algorithm(PIRG) has been 
proposed \cite{PIRG} as a new numerical algorithm for studying 
the ground state properties of strongly correlated fermion systems. 
The crucial point is that PIRG does not suffer from the 
negative sign problem and can be applied to any type of Hamiltonian 
in any dimension. The approximate ground state wavefunction is 
filtered out after numerical renormalization process in 
the path-integral formalism and expressed from the optimized linear 
combination of basis states in truncated Hilbert space. Since 
the explicit form of the wavefunction is obtained, 
the variational principle is satisfied and the method does not
lead to the sign problem in contrast to QMC. 
Compared to DMRG, the numerical renormalization is done to the
imaginary time direction irrespective of the spatial dimensionality of
the system.
Since the wavefunction is numerically given by using simple basis
representation, it makes it easier to compute physical properties. 

In Chapter \ref{chap1}, we will 
explain the whole ideas and procedures of
PIRG, such as renormalization group method, truncated Hilbert subspace
and extrapolation procedure, from the viewpoint of physical implications 
and compare PIRG with QMC and DMRG.
   
In Chapter \ref{chap2}, we will discuss the PIRG procedure from the
viewpoint of implementation.
There are some ideas and devices to make the PIRG calculation faster. We
discuss the computation time and necessary computer memory. We review
the structure of the PIRG procedure to discuss the parallelization of PIRG.  

In Chapter \ref{chap3}, we test PIRG efficiency by 
applying this method to the two-dimensional Hubbard model 
with nearest-neighbor transfers and compare the results with
those of exact diagonalization and QMC. 
We show that the PIRG gives the results with accuracy of around three digits for energy and around two digits for the momentum distributions and the spin correlations.

\section{Path-Integral Renormalization Group Algorithm}
\label{chap1}

\subsection{The whole procedures of PIRG}

PIRG method consists of the following procedures. 
\begin{itemize}
 \item Initial state.
       \begin{itemize}
	\item It is possible to use any kind of initial state. Generally, 
	      closer to the ground state, better as an initial state and it
	      is possible to use a linear combination of chosen basis
	      states $|\psi_{initial}\rangle=\sum_{i=1}^{L}c_{i}\slat{i}$
	      or to use a single basis state $|\phi\rangle$. 
       \end{itemize}
 \item Projection.
       \begin{itemize}
	\item The projection operator is introduced by $\exp\left(-\tau H\right)$ 
	      or by $1-\tau H$ to achieve the ground state. This projection
	      expands the stored Hilbert subspace and thus the number of
	      basis states increases.
       \end{itemize}
 \item Truncation.
       \begin{itemize}
	\item Generally, the projection makes the number of basis states 
	      increase larger than our computer memory, or even if 
	      they can be stored, it is impossible to deal with them in
	      the limited computer time. 
	      It is necessary to select important states and to keep the
	      dimension of the stored Hilbert subspace $L$. 
       \end{itemize}
 \item Iteration.
       \begin{itemize}
	\item By the projection process using proper projection operator
	      and the truncation process, the numerical renormalization
	      is achieved and the stored $L$ dimensional subspace approaches
	      the ground state. It is necessary to repeat this
	      renormalization process until the lowest energy of the truncated
	      Hilbert space converges to the lowest energy under the
	      condition that the dimension of subspace is $L$.
       \end{itemize} 
\end{itemize}

Empirically, to achieve the best approximate ground state under the allowed
dimension $L$, it is necessary to start PIRG using a good $L=1$ state.
In fact, the best result at $L=1$ represents nothing but the optimized
Hartree-Fock result. 
We start PIRG from $L=1$
and make $L$ larger step by step, using the previous converged state as
the initial state at next $L$-dimensional subspace. 
In addition to that, it
is important for a faster convergence to iterate the renormalization
process sufficiently at small $L$. 

\subsection{Renormalization group methods}

\subsubsection{Renormalization direction}

\begin{figure}[htbp]
 \epsfxsize=80mm
 \epsffile{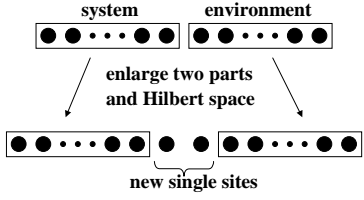}
 \caption{Graphical representation of the one step of the infinite-size
 system DMRG. 1D-system is divided into two parts, the system and the
 environment. Each of them are enlarged by one site and in this process
 the Hilbert space is also enlarged by the exactly represented one-site
 Hamiltonian. }\label{DMRG}
\end{figure}

The basic idea of the renormalization group methods is to keep the
relevant information of a system and integrate out the irrelevant one.
PIRG can be compared to the infinite-size system DMRG\cite{DMRG}. In the
infinite-size system DMRG, two single sites are inserted between the
system-part and the environment-part to enlarge these two parts as shown
in Fig.\ref{DMRG}. In this
process, using the exactly represented single-site Hamiltonians, the
stored states information increases temporarily before the truncation
process and the relevant information of the ground state is stored. 
This process is allowed only for one-dimensional system because the
dimension of the Hamiltonian matrix of two enlarged part, or the
dimension of the temporarily expanded Hilbert subspace, is too large to
treat for boundaries of multi-dimensional systems. 

In contrast to DMRG, the numerical renormalization is performed in the
imaginary time direction in PIRG. More concretely, the ground state
$|\psi_{g}\rangle$ is obtained by 
\begin{equation}
 |\psi_{g}\rangle=\lim_{\tau \to \infty} \exp[-\tau H]|\phi_{0}\rangle ,
\end{equation}
where $|\phi_{0}\rangle$ is an initial state. 
Although the finite temprature DMRG\cite{TDM1,TDM2,TDM3} has a
similarity to this renormalization in the imaginary time direction, it
can treat only the one-dimensional systems again because the transfer
matrix has to be utilized. 
Generally, the state $|\psi_{g}\rangle$ can be represented as a linear
combination of arbitrary basis states. However usually, 
the projection $\lim_{\tau \to \infty} \exp[-\tau H]$ can not be
performed in one operation. Following the Feynman's path-integral
formalism, 
by taking sufficiently small $\Delta\tau$, 
the projection procedure may, for example, be given as, 
\begin{eqnarray}  
 |\psi_{g}\rangle&=&\lim_{\Delta\tau\to 0}\lim_{n \to \infty} 
                           (\exp[-\Delta\tau H])^{n}
                                  |\phi_{0}\rangle . \label{project}
\end{eqnarray}
In quantum Monte Carlo methods, Eq.(\ref{project}) is usually 
implemented at a fixed large $n$ and sufficiently small $\Delta\tau$
which are enough to filter out the
ground state from the initial state $|\phi_{0}\rangle$.
In the PIRG method, as in DMRG approach, two identical new projection operators
$\exp[-\Delta\tau\hat{H}]$ are inserted between the stored states 
$|\psi_{s}\rangle$. Namely, 
one operates to the right state $|\psi_{s}\rangle$ while the other
operates to the left state $\langle\psi_{s}|$. 
When the stored state $|\psi_{s}\rangle$ is obtained after $n$
projection steps from the initial state $|\phi_{0}\rangle$, the
expectation value of a physical quantity
$\langle\hat{A}\rangle$ is represented by the initial state as
\begin{displaymath}
 \frac{\langle\psi_{s}|\hat{A}|\psi_{s}\rangle}
      {\langle\psi_{s}|\psi_{s}\rangle}
  \Leftarrow
 \frac{\langle\phi_{0}|\exp[-\Delta\tau \hat{H}]^{n} \hat{A}
        \exp[-\Delta\tau \hat{H}]^{n}|\psi_{0}\rangle}
      {\langle\phi_{0}|\exp[-\Delta\tau \hat{H}]^{n} 
         \exp[-\Delta\tau \hat{H}]^{n}|\psi_{0}\rangle} .
\end{displaymath}
Here, '$\Leftarrow$' indicates the truncation procedure in PIRG. 
After the step of the PIRG procedure, in other word, after the projection
operators $\exp[-\Delta\tau\hat{H}]$ are inserted, this expression is
changed into the following form.
\begin{eqnarray}
 \frac{\langle\psi_{s}|\exp[-\Delta\tau \hat{H}]\hat{A}
        \exp[-\Delta\tau \hat{H}]|\psi_{s}\rangle}
      {\langle\psi_{s}|\exp[-\Delta\tau \hat{H}]
         \exp[-\Delta\tau \hat{H}]|\psi_{s}\rangle}\qquad\nonumber\\
  \Leftarrow
 \frac{\langle\phi_{0}|\exp[-\Delta\tau \hat{H}]^{n+1} \quad \hat{A}
         \quad \exp[-\Delta\tau \hat{H}]^{n+1}|\psi_{0}\rangle}
      {\langle\phi_{0}|\exp[-\Delta\tau \hat{H}]^{n+1} \quad 
         \exp[-\Delta\tau \hat{H}]^{n+1}|\psi_{0}\rangle} . \nonumber
\end{eqnarray}
Because the dimension of the Hilbert subspace increases in this projection
process, the truncation process is necessary similarly to DMRG method. 
By this projection and truncation process, PIRG method achieves the
numerical renormalization in the imaginary time direction which is
represented in the path integral formalism.

\subsubsection{Path integral formalism for the Hubbard model in the Slater
  determinant basis} 
The formalism given below shares the similarity to that of the auxiliary
field Monte Carlo method\cite{QMC1}. In this paper, we use the
following Hubbard model Hamiltonian: 
\begin{eqnarray}
 H&=&H_{k}+H_{U},\nonumber\\
 H_{k}&=&-\sum_{<i,j>,\sigma}t_{ij}
  \left(c_{i\sigma}^{\dagger}c_{j\sigma}+h.c.\right),\nonumber\\
 H_{U}&=&U\sum_{i}
  \left(n_{i\uparrow}-\frac{1}{2}\right)
  \left(n_{i\downarrow}-\frac{1}{2}\right)\nonumber\\
 &=&U\sum_{i}n_{i\uparrow}n_{i\downarrow}-U\left(M-\frac{N}{4}\right),
  \label{simpleHamil}
\end{eqnarray}
where $i$ and $j$ represent the lattice points,
$c_{i\sigma}^{\dagger}\left(c_{i\sigma}\right)$ the creation
(annihilation) operator of an electron with spin $\sigma$ on the $i$-th
site, $n_{i\sigma}=c_{i\sigma}^{\dagger}c_{i\sigma}$, $t_{ij}$ the
transfer integral between the $i$-th site and the $j$-th site, $U$ the
on-site Coulomb interaction, $M=\sum_{i,\sigma}n_{i\sigma}$ and $N$
the number of the lattice sites. 

The projecting operator can be divided into the kinetic and the
interaction terms approximately.
\begin{equation}
 \exp\left[-\Delta\tau\left(\hat{H}_{k}+\hat{H}_{U}\right)\right]=
  \exp\left[-\Delta\tau\hat{H}_{k}\right] \exp\left[-\Delta\tau\hat{H}_{U}\right]
  + O\left(\left(\Delta\tau\right)^{2}\right).
\end{equation}

In this paper, we assume that the basis states are Slater
determinants. Hereafter, we use the notation $|\phi_{\sigma}\rangle$ to
represent a Slater determinant with spin $\sigma$ and $|\phi\rangle
=|\phi_{\uparrow}\rangle\otimes |\phi_{\downarrow}\rangle$. 
Since a Slater determinant is a single particle
state, the projection of single-body operator only changes a Slater
determinant to other single Slater determinant. Namely, 
\begin{equation}
 \exp\left[-\Delta\hat{H}_{k}\right]|\phi_{\sigma}\rangle 
  = |\phi_{\sigma}^{\prime}\rangle .
  \label{kineticPRO}
\end{equation}
Though the interaction term is a many-body term, the projection term of
it can be transformed into the sum of two single operator projections by
using the Stratonovich variable $s$:
\begin{eqnarray}
 \exp\left[-\Delta\tau U n_{m\uparrow}n_{m\downarrow}\right]
  \qquad\nonumber\\
  = \frac{1}{2}\sum_{s=\pm 1}
  \exp\left[\alpha (s)n_{m\uparrow}\right]
  \exp\left[\alpha (-s)n_{m\downarrow}\right], 
  \label{INT1}
\end{eqnarray}
where 
\begin{eqnarray}
 \alpha (s) = 2as-\frac{\Delta \tau U}{2} ,\label{INT2}\\
 a = \tanh^{-1}\sqrt{\tanh\left(\frac{\Delta\tau U}{4}\right)}.
  \label{INT3}
\end{eqnarray}
As a consequence, 
\begin{eqnarray}
 \exp\left[-\Delta\tau U n_{m\uparrow}n_{m\downarrow}\right]|\phi\rangle 
   &=& \frac{1}{2}\left(
		   |\phi^{m+}\rangle +|\phi^{m-}\rangle
		   \right)\label{INT4} \\
 \exp\left[-\Delta\tau\hat{H}_{U}\right]|\phi\rangle
  &=& \sum_{i=1}^{2^{N}} |\phi_{i}\rangle
\end{eqnarray}
In this way, the projection of the local interaction term, 
$\exp\left[-\Delta\tau Un_{m\uparrow}n_{m\downarrow}\right]$, changes a
Slater determinant to the sum of two Slater determinants. After
the projection of the $N$-sites interaction term $\exp\left[-\Delta\tau
\hat{H}_{U}\right]$, an original single Slater determinant
expands to the sum over $2^{N}$ Slater determinants. Though we
use Slater determinants and a projection $\exp[-\Delta\tau\hat{H}]$, 
a similar dimensional expansion occurs even if we use other kind
of basis states such as the site representation and a projection
$1-\hat{H}$. This is the reason why QMC sampling or PIRG truncation is
necessary.  

\subsection{Truncated Hilbert space}

\subsubsection{Variational principle}
The expectation value $\langle\rangle_{g}$ of a physical variable
$\hat{A}$ in the ground state $|\psi_{g}\rangle$ 
can be calculated with an arbitrary complete set of basis
states ${\slat{i}}$ as follows:
\begin{eqnarray}
 |\psi_{g}\rangle&=&\sum_{i}^{D_{Hilbert}}|\phi_{i}\rangle ,\nonumber\\
 \langle\hat{A}\rangle_{g}
  &=&\frac{\sum_{i,j}^{D_{Hilbert}}\lstat{i}\hat{A}\rstat{j}}
  {\sum_{i,j}^{D_{Hilbert}}\lstat{i}\phi_{j}\rangle} .
  \label{expectation}
\end{eqnarray}
However, the sum in the above equation is practically impossible in
general for easily available basis states because the number of them,
$D_{Hilbert}$, is usually comparable to the dimension of the whole
Hilbert space. 
QMC methods
deal with this problem by sampling the numerator and the denominator
separately:
\begin{equation}
  \langle\hat{A}\rangle\approx
   \frac{\sum_{(a,b)}^{N_{s}}
                \left( 
		      \overbrace
		      {\frac{\|\lstat{a}\phi_{b}\rangle\|}
		            {\sum_{i,j}^{D_{Hilbert}}\|\lstat{i}\phi_{j}\rangle\|}
		      }^{Probability}
		      \times
		      \overbrace
		      {\frac{\lstat{a}\phi_{b}\rangle}
		            {\|\lstat{a}\phi_{b}\rangle\|}
	              }^{Sign}
		      \times
		      \overbrace
		      {\frac{\lstat{a}\hat{A}\rstat{b}}
		            {\lstat{a}\phi_{b}\rangle}
                      }^{Sample}
		\right)}
	     {\sum_{(c,d)}^{N_{s}}
                \left( 
		      \underbrace
		      {\frac{\|\lstat{c}\phi_{d}\rangle\|}
		            {\sum_{i,j}^{D_{Hilbert}}\|\lstat{i}\phi_{j}\rangle\|}
		      }_{Probability}
		      \times
		      \underbrace
		      {\frac{\lstat{c}\phi_{d}\rangle}
		            {\|\lstat{c}\phi_{d}\rangle\|}
	              }_{Sign}
		\right)} \label{QMC}		
\end{equation}
where $\|\|$ represents the absolute value and $N_{s}$, the number of
Monte Carlo samples. 
Usually, these sampling processes are taken for an identical set 
$(c,d)=(a,b)$ in the Metropolis algorithm. 

Because the diagonal elements $\langle\phi_{a}|\phi_{a}\rangle$ are not
necessarily contained in Monte Carlo samples and
$\langle\phi_{a}|\phi_{b}\rangle$ in the sum are taken only
partially by sampling, QMC does not satisfy the variational principle
in the strict sense, although the deviation should be within the range of 
statistical error.
Several thousands samples are practically taken, 
to reach the accuracy with a relative error less than a few percents. 
In case of the system with the sign
problem, however, it is difficult to achieve the same accuracy 
by this sampling process.  

On the contrary, the variational principle is satisfied and the
sign problem is absent in PIRG because 
the ground state is represented approximately as a linear combination
of arbitrary basis states. 
\begin{equation}
 |\psi_{g}\rangle\approx 
  |\psi\rangle = \sum_{i=1}^{L}w_i|\phi_{i}\rangle . \label{F1}
\end{equation}
In Eq.(\ref{F1}), $\slat{i}$ are optimal basis states in the whole
Hilbert space. 
For example, orthogonal basis states such as site-represented ones
or non-orthogonal basis states such as Slater determinants can be used.
In any case, to take the sum in Eq.(\ref{expectation}) within the allowed
computation time, these basis states should be simple ones and the number of
them, $L$, be small. Of course, the relation between
$L$ and the truncation error may depend on the choice of basis states. 
Although we discuss the relation next, we do not know
it clearly now. Empirically, as we show the results in
Chapter \ref{chap3}, with the choice of the Slater determinant 
bases, limited and tractable basis states, such as hundreds basis
states, can reach the ground state with
the relative error less than a few percents for most of physical
quantities of our interest even when the Hilbert space is enormously
large. 

From theoretical and practical viewpoints, the relation between the error 
of estimated properties and the truncated Hilbert space dimensions $L$ is very
important. It is known that the relative error decreases exponentially
in DMRG\cite{DMRG}  when the number of states kept, or the dimension of the
subspace, increases. This makes DMRG powerful.
In DMRG, because the states are not treated explicitly, 
there is no restriction on the choice of bases. 
Namely, although both PIRG and DMRG are the methods to
make the truncated Hilbert space converge to the ground state, the
restrictions posed on the choice of basis states are different in two
methods. 

\subsubsection{The lowest energy in the truncated Hilbert space}

As shown in Eq.(\ref{F1}), the stored approximate ground state is
represented as, 
\begin{displaymath}
 \state = \sum_{i=1}^{L}w_i\slat{i}
\end{displaymath}
Because the basis states $|\phi_{i}\rangle$ which constitute the stored
Hilbert subspace are chosen so as to give the lowest energy within the
allowed dimensions of the Hilbert subspace, it
is necessary to optimize the coefficient $w_{i}$ in every truncation
process. The energy $E$ of the state $|\psi\rangle$
follows the equation, 
\begin{eqnarray}
 E=\frac{\sum_{i,j=1}^{L}[H]_{ij}w_{i}w_{j}}{\sum_{i,j=1}^{L}[F]_{ij}w_{i}w_{j}}
  \label{energy}
\end{eqnarray}
where,
\begin{eqnarray}
 \left[H\right]_{ij}=\langle\phi_{i}|\hat{H}|\phi_{j}\rangle \nonumber\\
 \left[F\right]_{ij}=\langle\phi_{i}|\phi_{j}\rangle .
  \label{defFH}
\end{eqnarray}

The lowest energy $E$ in this subspace is the lowest eigenvalue of this
subspace. Then, to obtain the lowest energy and its state, we solve the
following generalized eigenvalue problem. 
\begin{eqnarray}
  \sum_{j=1}^{L}[H]_{ij}w_{i}&=&E\sum_{j=1}^{L}[F]_{ij}w_{j}.
  \label{geneigen}
\end{eqnarray}

\subsection{Extrapolation on the dimension of the stored subspace to the
  dimension of the whole Hilbert space}

\subsubsection{Extrapolation of energy}
We need the extrapolation procedure to large $L$ to estimate the
systematic deviation, because the exact value is achieved only when $L$
becomes the dimension of the whole Hilbert space.
In practice it is difficult to analyze the relation between
the dimension $L$ and the systematic deviation of expectation values. 
In addition to that, the relation may depend on the choice of the
bases. Therefore, we do not use the argument $L$ in the extrapolation
function and introduce another extrapolation procedure. 
The important point is
that the extrapolation procedure gives more accurate estimate. 
We, however, note that after this extrapolation, the variational
principle may not necessarily be satisfied because the obtained energy
could be lower than the exact one within the extrapolation error. 
We define the difference between the ground state energy and the
expectation value in a given subspace as
\begin{equation}
 \delta E = \langle \hat{H}\rangle -\langle \hat{H}\rangle_{g}
\end{equation}
It can be shown\cite{Sorella} that the difference $\delta E$ vanishes 
linearly as a function of the energy variance $\Delta E$ defined by
\begin{equation}
 \Delta E=\frac{\langle\hat{H}^{2}\rangle - \langle\hat{H}\rangle^{2}}
  {\langle\hat{H}\rangle^{2}}
\end{equation}

We summarize the proof of this in the following way: 
First we put the approximate ground state $|\psi\rangle$ as
\begin{eqnarray}
 |\psi\rangle = c|\psi_{g}\rangle + d|\psi_{e}\rangle\\
 c^{2}+d^{2}=1\nonumber\label{statege}
\end{eqnarray}
where $|\psi_{g}\rangle$ and $|\psi_{e}\rangle$ are orthonormalized states.
We also define 
\begin{eqnarray}
 D_{1}\equiv\frac{\langle \hat{H}\rangle_{e}-\langle \hat{H}\rangle_{g}}
  {\langle \hat{H}\rangle_{g}}\quad ,\nonumber\\
 D_{2}\equiv\frac{\langle \hat{H}^{2}\rangle_{e}-\langle \hat{H}\rangle_{g}^{2}}
  {\langle \hat{H}\rangle_{g}^{2}}\quad ,\nonumber\\
 D_{3}\equiv\frac{\langle \hat{H}^{3}\rangle_{e}-\langle \hat{H}\rangle_{g}^{3}}
  {\langle \hat{H}\rangle_{g}^{3}}\quad ,
\end{eqnarray}
where
\begin{equation}
  \langle\hat{A}\rangle_{g}
  =\frac{\langle\psi_{g}|\hat{A}|\psi_{g}\rangle}{\langle\psi_{g}|\psi_{g}\rangle}
  ,\qquad
  \langle\hat{A}\rangle_{e}
  =\frac{\langle\psi_{e}|\hat{A}|\psi_{e}\rangle}{\langle\psi_{e}|\psi_{e}\rangle}.
  \nonumber
\end{equation}
In this notation, 
\begin{eqnarray}
 \langle\hat{H}\rangle &=& \langle\hat{H}\rangle_{g}
  +d^{2}\langle\hat{H}\rangle_{g}D_{1}\nonumber\\
 \langle\hat{H}^{2}\rangle &=& \langle\hat{H}\rangle_{g}^{2}
  +d^{2}\langle\hat{H}\rangle_{g}^{2}D_{2}\nonumber\\
 \langle\hat{H}^{3}\rangle &=& \langle\hat{H}\rangle_{g}^{3}
  +d^{2}\langle\hat{H}\rangle_{g}^{3}D_{3}\nonumber
\end{eqnarray}
\begin{eqnarray}
 \Rightarrow
  \left\{
   \begin{array}{lll}
    \delta E=d^{2}E_{0}D_{1}\\
    \Delta E=d^{2}\left(D_{2}-2D_{1}\right)+ d^{4}D_{1}\left(3D_{1}-2D_{2}\right) 
     + O\left(d^{6}\right)\nonumber
   \end{array}
    \right.  .\\  \label{de}
\end{eqnarray}
where $\langle\hat{A}\rangle = \langle\psi|\hat{A}|\psi\rangle
/\langle\psi|\psi\rangle$ and $E_{0}\equiv\langle\hat{H}\rangle_{g}$. 
When the stored state $|\psi\rangle$ is a good approximation of the
ground state, the coefficient $d$ is expected to be small. 
Then up to $O\left(d^{3}\right)$, 
\begin{equation}
 \delta E \propto \Delta E
\end{equation}
is satisfied.
This is the simplest extrapolation procedure we introduce.

We may introduce other series of extrapolation procedure. They are more time
consuming but can be more accurate.
As we discussed in the renormalization procedure, 
there are operators that can make a state closer to the ground
state. Here we take this kind of projection operator $\hat{R}$.
In this case, for more accurate extrapolation, $\delta E$ and
$\Delta E$ in the above simplest extrapolation procedure should be
calculated on the state $\hat{R}|\psi\rangle$. Namely, in the above
equations for any operators $\hat{A}$,
\begin{equation}
 \langle\hat{A}\rangle
 =\frac{\langle\psi|\hat{R}^{\dagger}\hat{A}\hat{R}|\psi\rangle}
 {\langle\psi|\hat{R}^{\dagger}\hat{R}|\psi\rangle}\label{last_step}
\end{equation}
may give a better estimate than that from
$\langle\psi|\hat{A}|\psi\rangle / \langle\psi|\psi\rangle$.
In the renormalization procedure, we have to consider the
restriction of the Hilbert subspace dimension due to computer
power. As shown in the above, however, if it is possible to calculate
in the form Eq.(\ref{last_step}) with the projection $\hat{R}$, it is
equivalent to calculate the physical value with the state
$\hat{R}|\psi\rangle$, which is in a larger Hilbert subspace than that 
the state $|\psi\rangle$ belongs.

As an example of the projection operator $\hat{R}$, we take a function
of Hamiltonian $\hat{H}$. When the absolute value of the ground state
energy is the largest in all the eigenvalues, operating the Hamiltonian
$\hat{H}$ makes a state closer to the ground state. From a viewpoint of
computation time, the number of the terms contained in the Hamiltonian of
the system without any long-range interaction is the order of $N$. If we
take the Hamiltonian
$\hat{H}$ as the projection operator $\hat{R}$, it is necessary to
calculate $\langle\hat{H}^{4}\rangle$ in $\Delta E$ but it is not
practical because the computation time increases to the order of 
$N^{4}$. 
Therefore in 
this paper we take $\sqrt{\hat{H}}$ as the projection operator.
To obtain the 
expectation value of $\hat{A}$ on the state $\hat{R}|\phi\rangle$, we
use Eq.(\ref{last_step}). In this case, Eq.(\ref{de})
is modified to the following.
\begin{eqnarray}
 \delta E\rightarrow\delta E_{sqrt}=\frac{\langle\hat{H}^{2}\rangle}
  {\langle\hat{H}\rangle}-\langle\hat{H}\rangle_{g} .\\
 \Delta E\rightarrow\Delta E_{sqrt}
  =\frac{\langle\hat{H}^{3}\rangle\langle\hat{H}\rangle 
  - \langle\hat{H}^{2}\rangle^{2}}{\langle\hat{H}^{2}\rangle^{2}} .
  \label{sqrtE}
\end{eqnarray}
The difference $\delta E_{sqrt}$ is always smaller than the difference
$\delta E$ and the extrapolation procedure can be done closer to the
ground state if we use $\delta E_{sqrt}$ and $\Delta E_{sqrt}$. One might
consider a problem whether the operator $\sqrt{\hat{H}}$ exists or not. 
In the representation of $\delta E_{sqrt}$ and $\Delta E_{sqrt}$,
however, $\sqrt{\hat{H}}$ does not appear but only $\hat{H}$ does and
the argument for $\delta E_{sqrt}$ and $\Delta E_{sqrt}$ remains
correct. Even if we ignore the projection operator
$\hat{R}=\sqrt{\hat{H}}$ in Eq.(\ref{last_step})
and only use Eq.(\ref{sqrtE}), the following equations are obtained. 
\begin{eqnarray}
 \Rightarrow
  \left\{
   \begin{array}{lll}
    \delta E_{sqrt}=d^{2}E_{0}\left(D_{2}-D_{1}\right)\nonumber\\
     \quad +d^{4}E_{0}D_{1}\left(2D_{1}-D_{2}\right)+O\left(d^{6}\right)\\
    \Delta E_{sqrt}
     =d^{2}\left(D_{3}+D_{1}-2D_{2}\right)\nonumber\\
    \quad 
     +d^{4}\left(D_{3}D_{1}-2D_{2}D_{3}-2D_{1}D_{2}+3D_{2}^{2}\right)
    +O\left(d^{6}\right)
   \end{array}
    \right. \\
\end{eqnarray}
By the same reason as that of the simple extrapolation procedure, the
above equation leads to 
\begin{equation}
 \delta E_{sqrt} \propto \Delta E_{sqrt}.
\end{equation}

Because both extrapolations are on the energy difference
and the energy variance, they can be plotted on the same parameter
plane. As shown in the above, however, two extrapolation plots are on
different lines. 
\begin{equation}
 \delta E \approx \frac{DE_{0}}{D_{2}-2D}\Delta E \label{eq1}
\end{equation}
\begin{equation}
 \delta E_{sqrt} \approx \frac{\left(D_{2}-D\right)E_{0}}{D_{3}+D-2D_{2}}
  \Delta E, \label{eq2}
\end{equation}
where the proportionality coefficients are different between Eq.(\ref{eq1})
and Eq.(\ref{eq2}). 
In this paper we use the $\delta E_{sqrt}-\Delta E_{sqrt}$
extrapolation to examine the accuracy of the ground state energy.

\subsubsection{Extrapolation of other physical values}

Because no commutation relation is expected between most operators
of the physical quantity $\hat{A}$ and Hamiltonian $\hat{H}$,
\begin{displaymath}
 \frac{\langle\phi|\sqrt{\hat{H}}\hat{A}\sqrt{\hat{H}}|\phi\rangle}
 {\langle\phi|\sqrt{\hat{H}}\sqrt{\hat{H}}|\phi\rangle}
 \ne
 \frac{\langle\phi|\hat{H}\hat{A}|\phi\rangle}
 {\langle\phi|\hat{H}|\phi\rangle}.
\end{displaymath}
Then we can not use the above second extrapolation procedure. From the same
reason as the computation time for obtaining the energy, higher order
projection is 
not practically possible within our accessible computer power. For the
moment, we do not know the suitable
projection operator $\hat{R}$. Then, we use
the simple extrapolation for other physical values than the energy. Some
modifications are, however, necessary
in the above discussion. The expectation value of $\hat{A}$ is
\begin{eqnarray}
 \langle\hat{A}\rangle =\langle\hat{A}\rangle_{g} 
  + d^{2}\langle\hat{A}\rangle_{g}D_{A} 
  + 2cd\langle\psi_{e}|\hat{A}|\psi_{g}\rangle ,
\label{otherPHYS}
\\
 D_{A}\equiv\frac{\langle\hat{A}\rangle-\langle\hat{A}\rangle_{g}}
  {\langle\hat{A}\rangle_{g}} .\nonumber
\end{eqnarray}
Different from the discussion on the extrapolation of the energy, in which
the energy difference and the energy variance is proportional to
$d^{2}$, the third term in $\langle\hat{A}\rangle$ is proportional
to $d$ and therefore, the extrapolation of the expectation value of
the operator $\hat{A}$ should be done on the square root of the energy
variance. 
Empirically, however, the $d$ term is so small that it is easy to do the
extrapolation on $\langle\hat{A}\rangle$ linearly to the energy variance
$\Delta E$, and we do so in this paper. Theoretically, it has been
proven\cite{Sorella} that if the operator $\hat{A}$ is a short-ranged
correlation function, $\langle\psi_{e}|\hat{A}|\psi_{g}\rangle$ in
Eq.(\ref{otherPHYS}) becomes zero in the 
thermodynamic limit as the following: Because $|\psi_{g}\rangle$ and
$|\psi_{e}\rangle$ are orthonormalized states, 
\begin{eqnarray}
 \left|\langle\psi_{g}|\hat{A}|\psi_{e}\rangle\right|^{2}=
  \left|\langle\psi_{g}|\hat{A}-\langle\hat{A}\rangle_{g}
      |\psi_{e}\rangle\right|^{2}\nonumber\\
 \leq \left\langle\left(\hat{A}-\langle\hat{A}\rangle_{g}\right)^{2}
        \right\rangle_{g}
\end{eqnarray}
where the latter inequality is led by Schwartz inequality and
$\langle\quad\rangle_{g}=\langle\psi_{g}|\quad|\psi_{g}\rangle$. 
When $\hat{A}$ is a short-ranged correlation function, the final term is
proportional to $1/N$ and becomes zero in the thermodynamic limit. 

These extrapolation procedures on energy and other physical values are
satisfied in a general ground state. They are not necessarily restricted
to the purpose of the application in PIRG method and not necessarily for
studying the ground state. It is possible to generalize these
extrapolation procedures for obtaining a more accurate eigenstate
properties from a series of approximations of eigenstate.  

\subsubsection{Comparison with previous methods}
Truncations of the Hilbert space were used in many numerical
method. 
Although DMRG is one of the most remarkable approaches, many other algorithms
applicable to higher-dimensional systems have also been proposed. 

Exact diagonalization method combined with the truncation of the Hilbert
space\cite{TrEx1,TrEx2,TrEx3} can treat lagrer systems than a normal
exact diagonalization method can do. 
The efficiency of the usage of basis state, however, is not sufficient
partly because the basis states are in site-representation. 
Therefore the convergence of the energy as a function of the dimension
of Hibert subspace is much worse than that of PIRG which uses Slater determinants
for basis states. 

On the other hand, the quantum Monte Carlo method is also combined with the
truncation of the Hilbert space and applied to fermion
systems\cite{MC_Tr1} and boson systems\cite{MC_Tr2}. 
Although in these methods, candidate basis states are generated by Monte
Carlo process of the whole Hamiltonian, in our experience, it is crucial
for the lowering of the energy to make the acceptance of them higher by
sequentially generating local, or one-site, projection. 
Renormalization and projection through the interaction term is achieved
more efficeintly by the local algorithmas we discuss in \S~\ref{chap2}.2.1.
Several devices to get out of local minima and to realize global minimum
are important in models of condensed matter systems.  

A large difference between the above numerical algorithm and PIRG is
the presece of the combined extrapolation procedure.
It can make errors smaller systematically.

\section{On the implementation of PIRG}
\label{chap2}

\subsection{Matrix representation}
The basic ideas of the following discussion are published~\cite{QMC1}.
At first, we explain the expression of states and expectation values. 
A Slater determinant $|\phi_{\sigma}\rangle$ is represented as an 
$N\times M$ matrix $[\phi_{\sigma}]$.
\begin{equation}
 |\phi_{\sigma}\rangle=\prod_{j=1}^{M}
  \left(\sum_{i=1}^{N}[\phi_{\sigma}]_{ij}c_{i\sigma}^{\dagger}\right)
  |0\rangle
\end{equation}
The inner product of two Slater determinants is
\begin{equation}
  \langle\phi_{a\sigma}|\phi_{b\sigma}\rangle 
    = \textrm{det}\left({}^{t}\left[\phi_{a\sigma}\right]
	  \left[\phi_{b\sigma}\right]\right).
	  \label{determinant}
\end{equation}
(See Appendix\ref{DET}). 
The matrix elements for the Hamiltonian and other operators 
are calculated from single-particle Green's function 
using the Wick's theorem, because a Slater determinant is a
single-particle state. A single-particle Green's function 
$[G_{\sigma}]$, which is represented as an $N\times N$ matrix, is
calculated by 
\begin{eqnarray}
 [G_{\sigma}^{ab}]_{ij}&=& 
  \frac{\langle\phi_{a\sigma}|
  c_{i\sigma}^{\dagger}c_{j\sigma}|\phi_{b\sigma}\rangle}
  {\langle\phi_{a\sigma}|\phi_{b\sigma}\rangle}\nonumber\\
 &=&\sum_{k=1}^{M}\sum_{l=1}^{M}\left[\phi_{b\sigma}\right]_{ik}
 \left[g_{\sigma}^{ab}\right]_{kl}\left[\phi_{a\sigma}\right]_{jl}
 \label{Green}
\end{eqnarray}
where  
\begin{equation}
 [g_{\sigma}^{ab}]=\left({}^{t}[\phi_{a\sigma}][\phi_{b\sigma}]\right)^{-1} .
\end{equation}
The procedure to use the Wick's theorem is explained in
the literature.\cite{QMC1}  
Since the stored state in PIRG is a linear combination of Slater
determinants, the expectation value of physical variables can be
calculated from one-particle Green's functions between all the stored
Slater determinants by using the Wick's theorem. 

Next we explain the projection processes following the
above representation. 
The kinetic term projection, Eq.(\ref{kineticPRO}), is expressed by using 
an $N\times N$ matrices $[M_{0}]$ and $[K]$ in the following way:
\begin{equation}
 [\phi_{\sigma}^{\prime}]_{ij}=\sum_{k=1}^{N}[M_{0}]_{ik}[\phi_{\sigma}]_{kj}
  \qquad\textrm{for }\sigma=\uparrow,\downarrow 
\end{equation}
where 
\begin{displaymath}
 [M_{0}]=\exp[K]
\end{displaymath}
\begin{equation}
 [K]_{ij}=-\Delta \tau t_{ij}.
\end{equation}
From Eq.(\ref{INT1}) the local-interaction term projection,
Eq.(\ref{INT4}), is expressed as the following: 
\begin{eqnarray}
 |\phi^{m+}\rangle&=&
    \left(\sum_{k=1}^{N}\left[M_{1}\left(1\right)\right]_{ik}
                      \left[\phi_{\uparrow}\right]_{kj}\right)
  \otimes
  \left(\sum_{k=1}^{N}\left[M_{1}\left(-1\right)\right]_{ik}
                      \left[\phi_{\downarrow}\right]_{kj}\right)\nonumber\\
 |\phi^{m-}\rangle&=&
    \left(\sum_{k=1}^{N}\left[M_{1}\left(-1\right)\right]_{ik}
                      \left[\phi_{\uparrow}^{m-1}\right]_{kj}\right)
  \otimes
  \left(\sum_{k=1}^{N}\left[M_{1}\left(1\right)\right]_{ik}
                      \left[\phi_{\downarrow}\right]_{kj}\right),
		      \label{matM1}
\end{eqnarray}
where $M_{1}$ is an $N\times N$ matrix,
\begin{eqnarray}
 [M_{1}\left(s\right)]_{ij}=
  \left\{
   \begin{array}{lll}
     \exp\left[2as-
	  \frac{\displaystyle{\Delta\tau U}}{\displaystyle{2}}\right]\qquad 
     \textrm{for } i=j=m \\
    1\qquad\textrm{       for }i=j,\quad i\ne m\\
    0\qquad\textrm{       otherwise}
   \end{array}
    \right. .\label{M1}
\end{eqnarray}

\subsection{Computation time}

\subsubsection{Whole procedures of the renormalization group method}
Empirically we find that it is better to perform the projection and
truncation at the projection of every local-interaction term 
$\exp\left[-\Delta\tau Un_{m\uparrow}n_{m\downarrow}\right]$. 
Hence we employ this {\it local algorithm}. 
Using this local algorithm, we summarize below how one PIRG iteration step 
$\exp\left[-\Delta\tau\hat{H}\right]
\left[\sum_{a=1}^{L}c_{a}|\phi_{a}\rangle\right]$ proceeds. 
Note that in this paper $|\phi_{a}\rangle$ represents
$|\phi_{a\uparrow}\rangle\otimes |\phi_{a\downarrow}\rangle$, 
namely a direct product of Slater determinants for each spin. 
\begin{itemize}
 \item Choose one basis state.
       \begin{itemize}
	\item
	     Choose $|\phi_{a}\rangle$ from $L$ stored basis states 
	     $\left\{|\phi_{1}\rangle,|\phi_{2}\rangle,
	      \ldots,|\phi_{L}\rangle\right\}$, which will be operated by 
          $\exp\left[-\Delta\tau\hat{H}\right]$
       \end{itemize}
 \item Projection by $\exp\left[-\Delta\tau\hat{H}_{k}\right]$.\\   
       $\Rightarrow\quad computation\quad time\propto
       \left(LN^{3}+L^{3}\right)$
       \begin{itemize}
	\item Perform 
	      $|\phi_{a}^{\prime}\rangle =\exp\left[-\Delta\tau\hat{H}_{k}\right]
	      |\phi_{a}\rangle$ following 
	      Eq.(\ref{kineticPRO}) and calculate the inner products
	      and Hamiltonian elements between
	      \begin{displaymath}
	       |\phi_{a}^{\prime}\rangle
	      \end{displaymath}
	      and 
	      \begin{displaymath}
	       \left\{|\phi_{1}\rangle,|\phi_{2}\rangle, \ldots,|\phi_{a-1}\rangle,
	      |\phi_{a+1}\rangle,\ldots,|\phi_{L}\rangle\right\} .
	      \end{displaymath}
	      $\rightarrow\qquad computation\quad time\propto LN^{3}$   
	\item Truncation is performed by comparing the lowest energies 
	      obtained from Eq.(\ref{geneigen}) in two subspaces which
	      consist of the following two sets of $L$ basis states:  
	      \begin{eqnarray}
		\left\{
		 \begin{array}{lll}
		  \left\{|\phi_{1}\rangle,|\phi_{2}\rangle,
			  \ldots,|\phi_{a-1}\rangle,|\phi_{a}\rangle,
			  |\phi_{a+1}\rangle,\ldots,|\phi_{L}\rangle\right\}\\
		  \left\{|\phi_{1}\rangle,|\phi_{2}\rangle,
			  \ldots,|\phi_{a-1}\rangle,|\phi_{a}^{\prime}\rangle,
			  |\phi_{a+1}\rangle,\ldots,|\phi_{L}\rangle\right\}\\
		 \end{array}
                \right.\nonumber
	      \end{eqnarray}
	      and by employing one of these two basis states set which
	      gives the lower energy. In other words, take
	      $|\phi_{a}\rangle$ or $|\phi_{a}^{\prime}\rangle$ 
	      to be a next basis state $|\phi_{a}^{0}\rangle$.\\
	      $\rightarrow\qquad computation\quad time\propto L^{3}$   
       \end{itemize}
 \item Projection by $\exp\left[-\Delta\tau\hat{H}_{U}\right]$.\\   
       $\Rightarrow\quad computation\quad time\propto
       \left(LN^{3}+L^{3}N\right)$ 
       \begin{itemize}
	\item Perform 
	      \begin{displaymath}
	      \frac{1}{2}\left(|\phi_{a}^{m+}\rangle + |\phi_{a}^{m-}\rangle
	       \right)
	      =\exp\left[-\Delta\tau Un_{m\uparrow}n_{m\downarrow}\right]
	      |\phi_{a}^{m-1}\rangle
	      \end{displaymath}
	      following Eq.(\ref{INT4}) and calculate the inner
	      products and Hamiltonian elements between
	      \begin{displaymath}
	       |\phi_{a}^{m+}\rangle\qquad\textrm{or}\qquad|\phi_{a}^{m-}\rangle
	      \end{displaymath}
	      and
	      \begin{displaymath}
	       \left\{|\phi_{1}\rangle,|\phi_{2}\rangle,
	       \ldots,|\phi_{a-1}\rangle,
	       |\phi_{a+1}\rangle,\ldots,|\phi_{L}\rangle\right\} .
	      \end{displaymath}
	       $\rightarrow\qquad computaiton\quad time\propto LN^{2}$   
	\item Truncation is performed by comparing the lowest energies 
	      obtained from Eq.(\ref{geneigen}) in three subspaces which
	      consist of the following three sets of $L$ basis states:  
	      \begin{eqnarray}
		\left\{
		 \begin{array}{lll}
		  \left\{|\phi_{1}\rangle,|\phi_{2}\rangle,
		   \ldots,|\phi_{a-1}\rangle,|\phi_{a}^{m-1}\rangle,
		   |\phi_{a+1}\rangle,\ldots,|\phi_{L}\rangle\right\}\\
		  \left\{|\phi_{1}\rangle,|\phi_{2}\rangle,
		   \ldots,|\phi_{a-1}\rangle,|\phi_{a}^{m+}\rangle,
		   |\phi_{a+1}\rangle,\ldots,|\phi_{L}\rangle\right\}\\
		  \left\{|\phi_{1}\rangle,|\phi_{2}\rangle,
		   \ldots,|\phi_{a-1}\rangle,|\phi_{a}^{m-}\rangle,
		   |\phi_{a+1}\rangle,\ldots,|\phi_{L}\rangle\right\}\\
		 \end{array}
                \right.\nonumber
	      \end{eqnarray}
	      and employing one of these three basis states set which gives the
	      lowest energy. 
	      The chosen basis state $|\phi_{a}^{m-1}\rangle$, 
	      $|\phi_{a}^{m+}\rangle\textrm{ or}
	      |\phi_{a}^{m-}\rangle$ is taken to be a next basis state
	      $|\phi_{a}^{m}\rangle$. \\ 
	      $\rightarrow\qquad computation\quad time\propto L^{3}$   
	\item Repeat the above local interaction projection for all the sites 
	      $m=1,2,\ldots,N$.
       \end{itemize}
 \item Repeat the above
       $\exp\left[-\Delta\tau\hat{H}\right]|\phi_{a}\rangle$ projection
       for all basis states $a=1,2,\ldots,L$.
\end{itemize}
\begin{equation}
 \Longrightarrow\quad The\quad total\quad computation\quad time\propto 
  L^{2}N^{3}+L^{4}N\label{time}   
\end{equation}

Though the basic computation time of PIRG is listed above, an 
efficient convergence procedure is important for reducing the
computation time. 
There may appear many states with local minima structure in energies
in the PIRG convergence process. Even if the convergence is not perfect,
which is actually the case in most of our experience, the extrapolation
procedure can be performed for the ground state properties. 
However, the worse converged state gives the value with larger
error. 
Because the extrapolation itself using the energy variance is the
formalism to obtain 
values of an eigenstate of the Hamiltonian, in case of worse convergence
than a limit, the extrapolation procedure could give the value not for
the ground state but for an excited state. Therefore it is important to
improve PIRG convergence 
process by a combination with some existing methods to avoid occurrence
of trapping in local
minima. At present, we have empirically learned that the sufficient
convergence at small 
subspace is crucial at the early stage of the process of extending $L$. 
In this paper, we have numerically realized
convergence to the state of the Hartree-Fock
approximation at $L=1$ and iterated hundreds projections at 
small $L$, for example for $L$ less than $10$. However 
more systematic methods are desired and are left for future studies. 

\subsubsection{Details on reduction of computation time}
Since inner products and Hamiltonian elements are calculated from
determinants and inverse matrices, the computation time for them is
usually proportional to $N^{3}$. In the process of local-interaction
projection, however, it can be reduced to $N^{2}$.  

Here, to explain the procedure to reduce the computation time from
$N^{3}$ to $N^{2}$, we simply consider the state
$\sum_{k=1}^{N}\left[M_{1}\left(s\right)\right]_{ik}
\left[\phi_{a\sigma}\right]_{kj}$ in Eq.(\ref{matM1}) for the following
discussion. When the basis state of the $a$-th Slater determinant with
spin $\sigma$ is updated by the projection of the $m$-th site local
interaction as, 
\begin{equation}
 \left[\phi_{a\sigma}\right]_{kj}\Longrightarrow
  \sum_{k=1}^{N}\left[M_{1}\left(s\right)\right]_{ik}
  \left[\phi_{a\sigma}\right]_{kj} 
\end{equation}
the inner products change as,
\begin{eqnarray}
 \langle\phi_{a\sigma}|\phi_{b\sigma}\rangle\Rightarrow
  \left\{
   \begin{array}{lll}
    \left(1+\delta\left(s\right)\left[G_{\sigma}^{ab}\right]_{mm}\right)\times 
     \langle\phi_{a\sigma}|\phi_{b\sigma}\rangle\quad\textrm{for }b\ne a\\
    \left(1+\delta\left(s\right)
     \left[\tilde{G}_{\sigma}^{ab}\right]_{mm}\right)\times 
     \langle\phi_{a\sigma}|\phi_{b\sigma}\rangle\quad\textrm{for }b=a\\
   \end{array}
  \right.
\end{eqnarray}
and the Green's functions change as, 
\begin{eqnarray}
 \left[G_{\sigma}^{ab}\right]_{ij}
 \Rightarrow
  \left\{
   \begin{array}{lll}
    \left(\delta\left(s\right)\delta_{im}+1\right)
     \left(\left[G_{\sigma}^{ab}\right]_{ij}
      -\frac{\displaystyle{\left[G_{\sigma}^{ab}\right]_{im}\delta\left(s\right)
				\left[G_{\sigma}^{ab}\right]_{mj}}}
	{\displaystyle{1+\left[G_{\sigma}^{ab}\right]_{mm}\delta\left(s\right)}}
    \right)\quad\textrm{for }b\ne a\\
    \left(\delta\left(s\right)\delta_{mj}+1\right)
     \left(\left[\tilde{G}_{\sigma}^{ab}\right]_{ij}
      -\frac{\displaystyle{\left[\tilde{G}_{\sigma}\right]_{im}
      \delta\left(s\right)
      \left[\tilde{G}_{\sigma}\right]_{mj}}}
	{\displaystyle{1+\left[\tilde{G}_{\sigma}\right]_{mm}\delta\left(s\right)}}
    \right)\quad\textrm{for }b=a\\
   \end{array}
  \right.
\end{eqnarray}
where
\begin{eqnarray}
 \left[\tilde{G}_{\sigma}\right]_{ij}&=&
    \left(\delta\left(s\right)\delta_{im}+1\right)
     \left(\left[G_{\sigma}^{ab}\right]_{ij}
      -\frac{\displaystyle{\left[G_{\sigma}^{ab}\right]_{im}\delta\left(s\right)
				\left[G_{\sigma}^{ab}\right]_{mj}}}
	{\displaystyle{1+\left[G_{\sigma}^{ab}\right]\delta\left(s\right)}}
    \right)\\
     \delta\left(s\right)&=&\exp\left[2as-
	  \frac{\displaystyle{\Delta\tau U}}{\displaystyle{2}}\right]-1.
\end{eqnarray}
Here $i,j$ is from $1$ to $N$ and $\delta_{im}$ is the Kronecker's
$\delta$ symbol.
In this way the computation 
time for local interaction projection is reduced to that proportional to
$N^{2}$. The similar reduction in the computation time is explained in
detail in the literature\cite{QMC1}.
  
\subsubsection{Devices on extrapolation procedure}

For the extrapolation process after the PIRG convergence, it is
necessary to calculate $\langle\hat{H}\rangle$,
$\langle\hat{H}^{2}\rangle$, $\langle\hat{H}^{3}\rangle$ and expectation
values of other operators. Because the Hamiltonian does not have any
long-range interaction term, the computation time for
$\langle\hat{H}^{n}\rangle$ is proportional to $N^{n}$ except for the
computation time for single particle Green's function $N^{3}$. The
coefficient of $N^{n}$ is, however, large and it is useful to decrease this
computation time by the following procedure. 

Because a converged state by PIRG is a linear combination of Slater
determinants and the kinetic term projection
$\exp[-\Delta\tau\hat{H}_{k}]$ does not increase the number of Slater
determinants, it is easier to calculate the right hand side of
the following equation than to calculate the left hand side 
from single particle Green's functions in the following equation: 
\begin{equation}
 \frac{\langle\psi|\hat{A}\hat{H}_{k}|\psi\rangle}
  {\langle\psi|\psi\rangle}=
  \frac{1}{\Delta\tau}\left(\frac{\langle\psi|\hat{A}|\psi\rangle}
		       {\langle\psi|\psi\rangle}
     - \frac{\langle\psi|\hat{A}
     \exp\left[-\Delta\tau\hat{H}_{k}\right]|\psi\rangle}
     {\langle\psi|\psi\rangle}\right)
  +O\left(\Delta\tau\right),\label{zuru}
\end{equation}
where we take sufficiently small $\Delta\tau$.
In this way, the computation time of the term, such as
$\hat{H}^{3}_{k}$, can be reduced.

\subsection{Required memory}

The largest share of the memory is exhausted for the storage of the
elements in all the elements of the Slater determinants in the basis
states and also for that of the Green's function. Here we assume scalar 
information takes $8byte$ and the number of basis states $L$ is $500$ to
roughly estimate the required maximum memory size. 
Slater determinant is an $N\times M$
matrix for each spin, and then near half filling of the Hubbard model,
this is comparable to $N^{2}$ scalar 
elements. As we see in (\ref{Green}), Green's function is represented by an $N\times N$ matrix for
each spin, and then this contains $2N^{2}$ scalars. If we store all the
Green's function data among the $L$ Slater determinants, about $83Gbyte$
is necessary for $12\times 12$ system and it cannot be easily stored in our
available computer. For this reason, we store only the  Green's function
between the Slater determinant just on the
operation of  $\exp[-\Delta\tau\hat{H}]$ and the others.   
Then the necessary memory for the state and Green's function is the
order of 
$3LN^{2}\times 8byte$, which is about $250Mbyte$ for $12\times 12$ system.
 
\subsection{Parallelization}
Parallelization of a code is a promising way to improve the performance
of the computation by dividing a large set of calculations into
several smaller pieces and execute them independently of each other.
We have tried to parallelize the code on the distributed memory system
using MPI. As shown in Eq.(\ref{time}), the computation time for a
single slice projection $\exp[-\Delta\tau\hat{H}]$ is
proportional to $L\times \left(LN^{3}+L^{3}N\right)$. 
Because the first factor $L$ refers to the iteration on all the basis
states which are related to each other in the evaluation of the energy,
it is difficult to parallelize this process. 
Then we consider the parallelization of the part $LN^{3}+L^{3}N$.  

The first term $LN^{3}$ is related to the calculation of inner products
and Green's functions between a basis state on the operation and the other
basis states. Each calculation is independent and it can be
parallelized. For this parallelization, the memory for each state and
each Green's function have to be distributed over each processor. For
example, the $1$-st state and the $1$-st Green's function are on the
$1$-st processor memory, and the $2$-nd state and the $2$-nd Green's
function are on the $2$-nd processor memory, etc..

The second term $L^{3}N$ is related to the iteration of the calculation
of the lowest energy of the stored Hilbert subspace by
Eq.(\ref{geneigen}) for each local projection. The matrix related to
this generalized eigenvalue problem is $L\times L$. In practice, $L$ is
the order of hundreds and it is difficult to parallelize efficiently the
eigenvalue problem of such a small matrix. 
However by performing the projection $\exp[-\Delta\tau\hat{H}]$ 
for some different choices of $\Delta\tau$ in parallel and employing the
result which gives the lowest energy among the choices of $\Delta\tau$,  
the convergence becomes faster and computation time for
Eq.(\ref{geneigen}) does not increase. In this way, we can reduce the total
computation time for convergence by parallelization.

\section{Evaluation of PIRG on Hubbard model}
\label{chap3}

\subsection{Model}
We apply PIRG to 
the Hubbard model Hamiltonian (\ref{simpleHamil}) on a two-dimensional
square lattice with nearest-neighbor transfers $t$. 
\begin{eqnarray}
 t_{ij}=
  \left\{
   \begin{array}{lll}
    t=1.0 \textrm{ if $(i,j)$ are the nearest neighbor sites}\\
    0 \textrm{    otherwise}
   \end{array}
    \right. \label{noFrust}
\end{eqnarray}
We take $t=1.0$ as the energy scale. 
Since these models have particle-hole symmetry at half filling,
the quantum Monte Carlo does not suffer from the sign problem at half
filling. By comparing with the QMC and exact
diagonalization results in the literature, 
we show PIRG results on the energy, the momentum distribution and the
equal-time spin correlations and discuss the accuracy and efficiency of
PIRG on these models. 

\subsection{Comparison of energy}

\subsubsection{PIRG result}

\begin{figure}
 \epsfxsize=80mm
 \epsffile{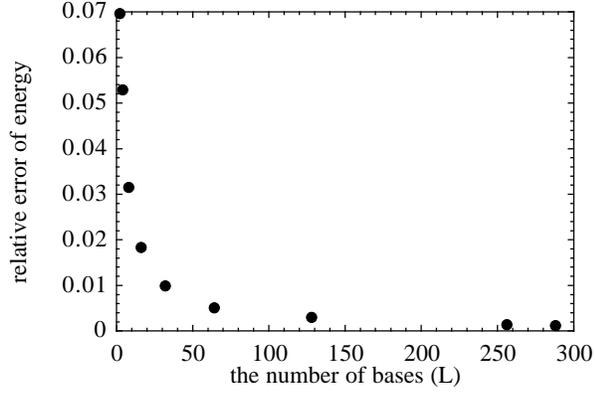}
 \caption{Relative error $\delta E/|E|$ in the ground state energy
 for the $6\times2$ Hubbard model with 5 up 5 down electrons and the
 fully periodic boundary condition at $U/t=4.0$.  The reference
 ground state energy is estimated from the exact diagonalization.}
 \label{L_re62}
\end{figure}

\begin{figure}
 \epsfxsize=80mm
 \epsffile{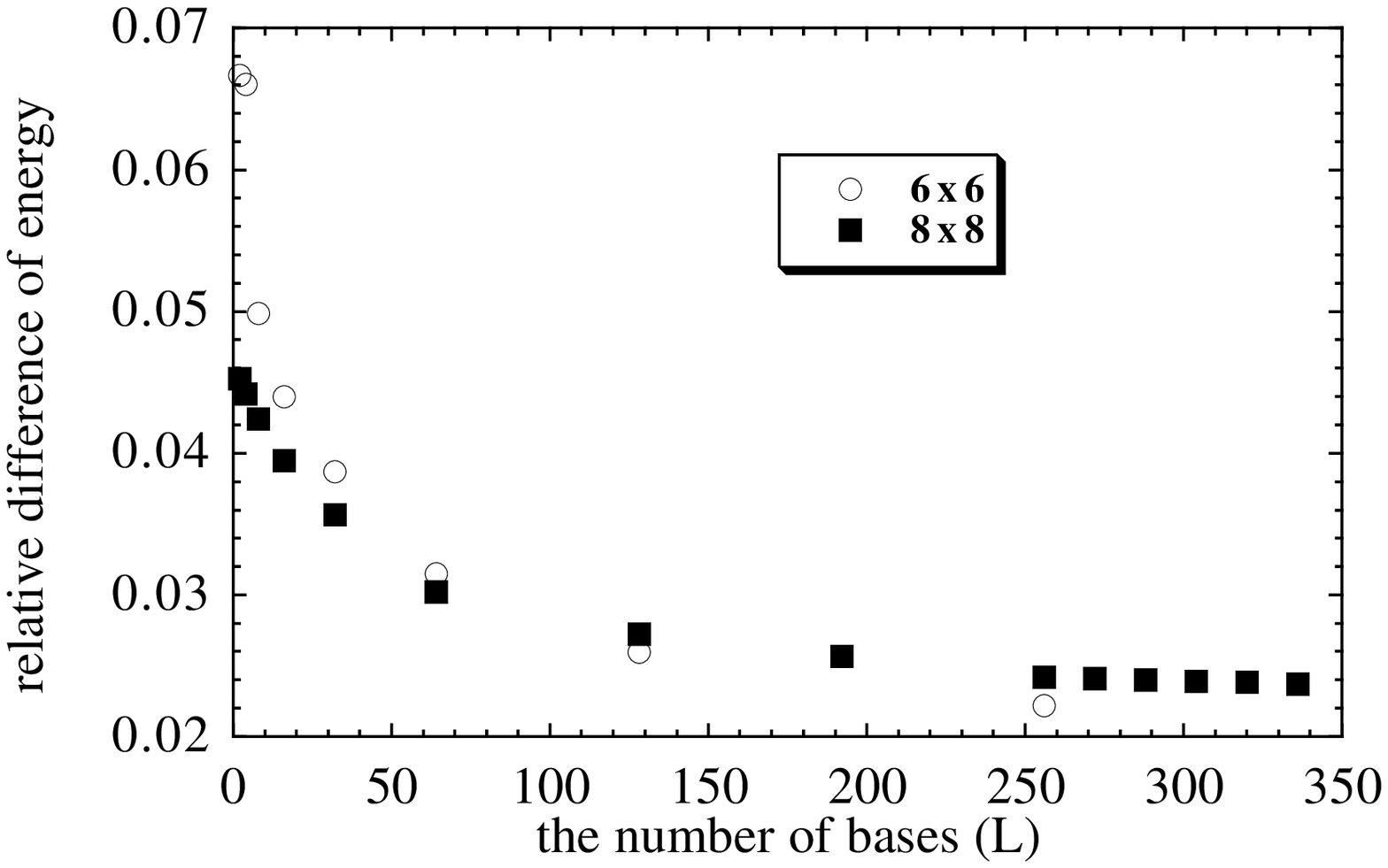}
 \caption{Relative difference $\delta E/|E|$ in the ground state energy
 for the Hubbard models at half filling
 with the fully periodic boundary condition at $U/t=4.0$. The reference
 ground state energy is estimated from QMC\cite{QMC2}.}
 \label{L_rehol}
\end{figure}

\begin{figure}
 \epsfxsize=80mm
 \epsffile{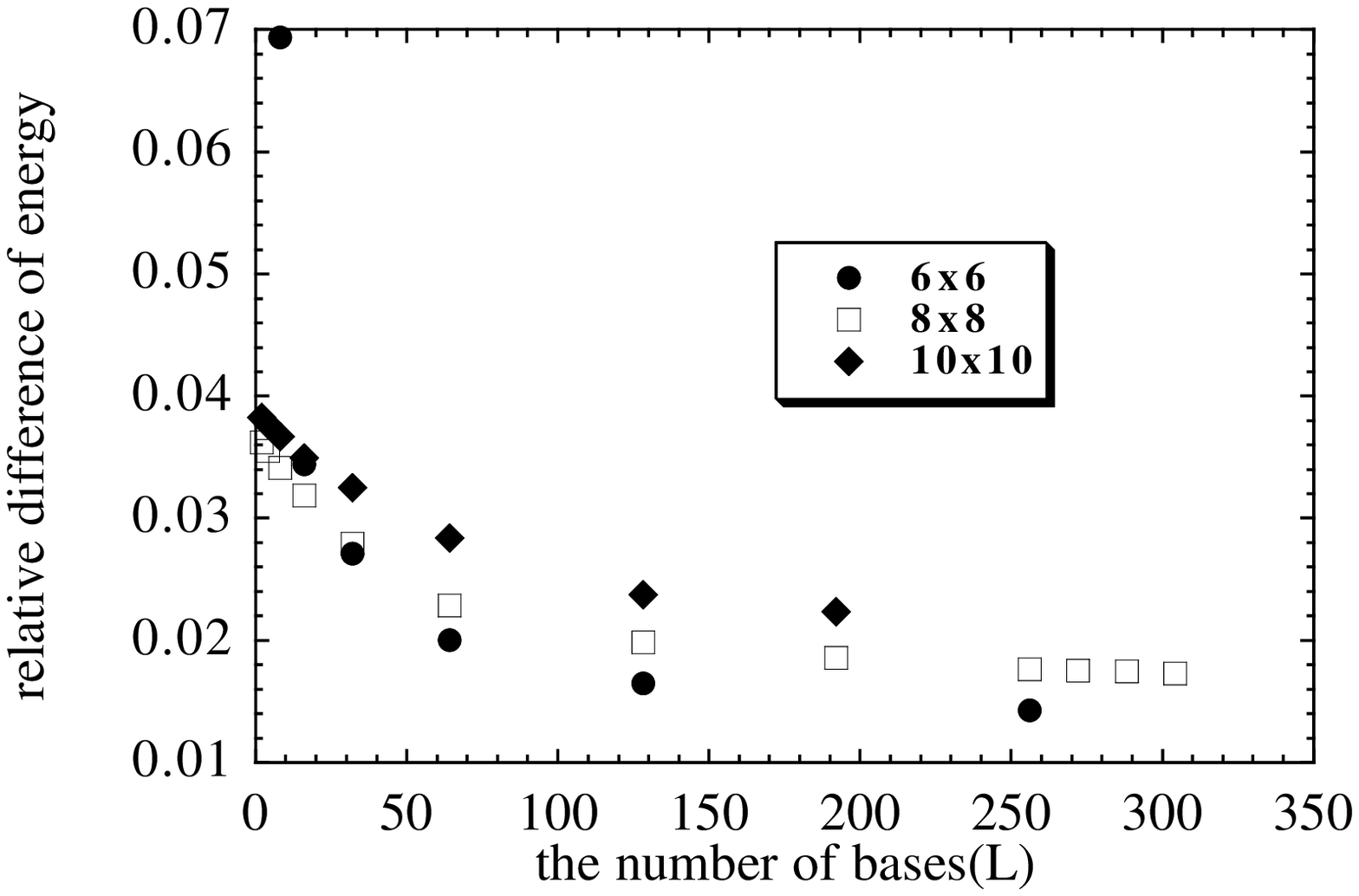}
 \caption{Relative difference $\delta E/|E|$ in the ground state energy
 for the Hubbard models with two hole doped from half filling and fully
 periodic boundary condition at $U/t=4.0$. The reference ground state
 energy is estimated from QMC\cite{QMC2}.}
 \label{L_rehal}
\end{figure}

First, we show the converged results by PIRG before the extrapolation
procedure and discuss the relative error. In $\S$\ref{chap1} we
refer to the DMRG method in which the
relative error decreases exponentially as a function of the dimension of
the stored Hilbert subspace. Because there is a restriction on the
choice of basis in PIRG, the relative error dependence on the dimension
of the stored Hilbert subspace is different between DMRG and
PIRG. Figure \ref{L_re62} shows the relative error and
Figs. \ref{L_rehol} and \ref{L_rehal} show the relative
difference between PIRG and QMC results. In case of a small system such
as $6\times 2$ Hubbard model, the relative error is smaller than 1
percent. For larger systems near half filling, the 
relative error is larger but less than a few percent independent of the
system size. This feature holds both at half filling and near half
filling. 
Note that the results at $L=1$ correspond to the Hartree-Fock
estimates. 


\subsubsection{Extrapolation results}
\begin{figure}
 \epsfxsize=80mm
 \epsffile{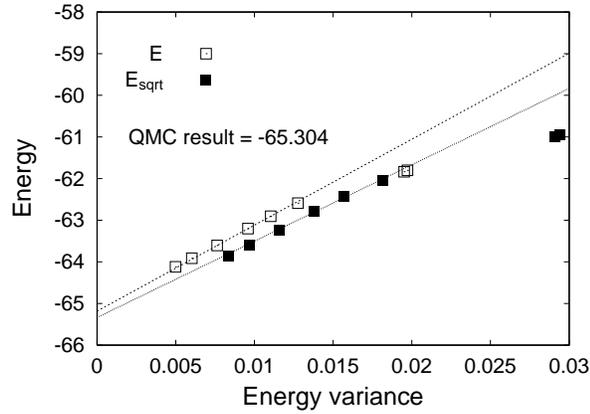}
 \caption{Extrapolation of the energy to the zero energy variance for a
 $6\times 6$ Hubbard model, 17 up 17 down electrons with the fully periodic boundary condition at
 $U/t=4.0$}
 \label{ex1}
\end{figure}

\begin{figure}
 \epsfxsize=80mm
 \epsffile{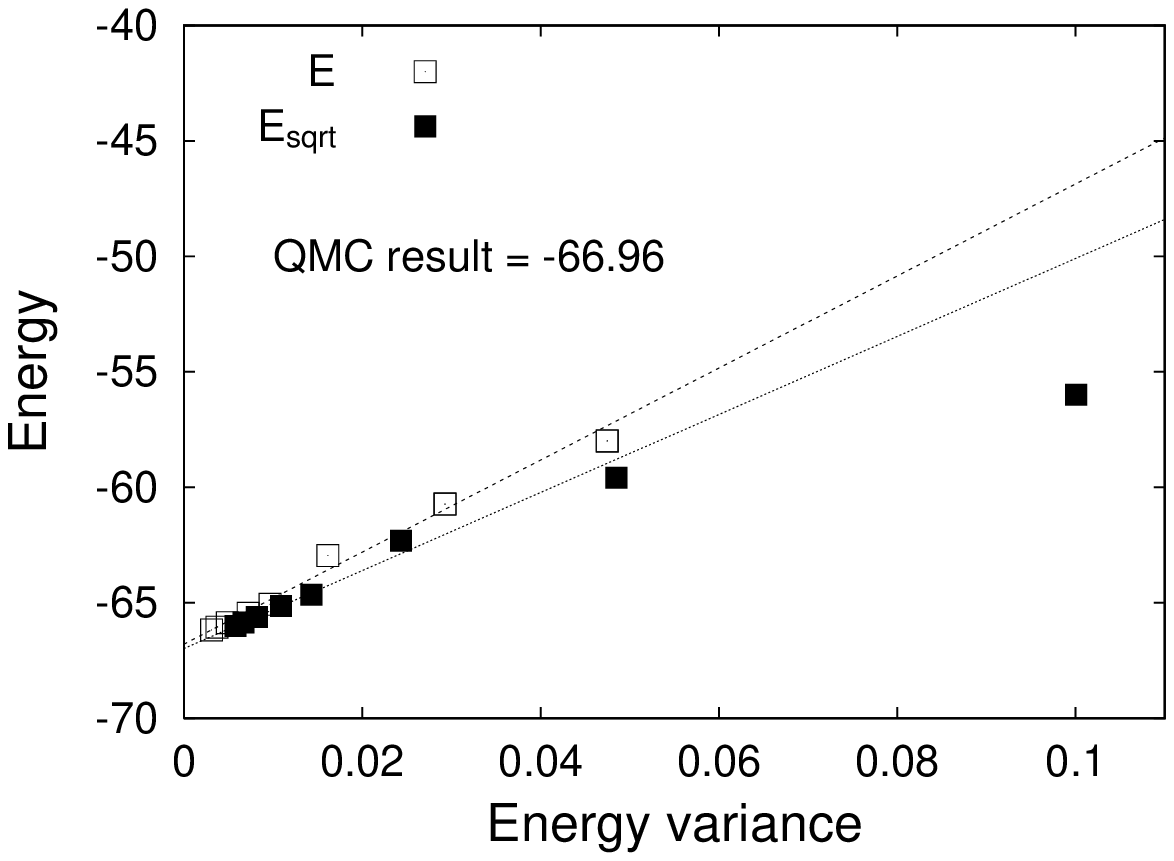}
 \caption{Extrapolation of the energy to the zero energy variance for a
 $6\times 6$ Hubbard model, 18 up 18 down electrons with the fully periodic boundary condition at
 $U/t=4.0$}
 \label{ex2}
\end{figure}

The above results are improved to more accurate estimates by the
extrapolation procedure. Here we show the simple extrapolation results
$E$ with the state $|\phi\rangle$ and the extrapolation results
$E_{sqrt}$ with the state $\sqrt{\hat{H}}|\phi\rangle$ for the same
$6\times 6$ Hubbard model by taking $L$ up to $256$. 
Both extrapolation procedures for two models
are shown in Figs. \ref{ex1} and \ref{ex2}. Two extrapolated results
should meet at the same value in principle, although there is a small
difference, which may be thought to be PIRG error. Because we do not
know which one is 
closer to the exact value and empirically it is better for the linear
function fitting to use the value obtained from the state
$\sqrt{\hat{H}}|\phi\rangle$, hereafter we employ the extrapolation
using $\sqrt{\hat{H}}|\phi\rangle$. The relative errors and
differences after the extrapolation are shown in Table \ref{Eerror}.
For most of the systems, PIRG can give results of the ground state energy
with less than 0.3 percent relative error or difference from QMC results.


\begin{table}
 \caption{relative error and difference of the ground state energy after
 extrapolation.}
 \begin{tabular}{@{\hspace{\tabcolsep}\extracolsep{\fill}}cccc}
  \hline 
  system & PIRG results & exact diagonalization and Monte Carlo results & 
  relative difference or error \\
  \hline
  $6\times 2$,  $5\uparrow,5\downarrow$ 
           & -25.6999 $\pm$ 0.0005 & -25.6952 & 0.00018\\ 
  $6\times 6$,  $17\uparrow,17\downarrow$
           & -65.12 $\pm$ 0.02 & -65.30 $\pm$ 0.04 & 0.0028\\
  $6\times 6$,  $18\uparrow,18\downarrow$
           & -66.92 $\pm$ 0.04 & -66.96 $\pm$ 0.07 & 0.00060\\ 

  $8\times 8$,  $31\uparrow,31\downarrow$
           & -117.8 $\pm$ 0.1 & -117.70 $\pm$ 0.06 & 0.00085\\ 
  $8\times 8$,  $32\uparrow,32\downarrow$
           & -119.4 $\pm$ 0.1 & -119.23 $\pm$ 0.06 & 0.0014\\ 
  \hline
 \end{tabular}
 \label{Eerror}
\end{table}

\subsection{Comparison for other physical quantities}

\begin{figure}
 \epsfxsize=80mm
 \epsffile{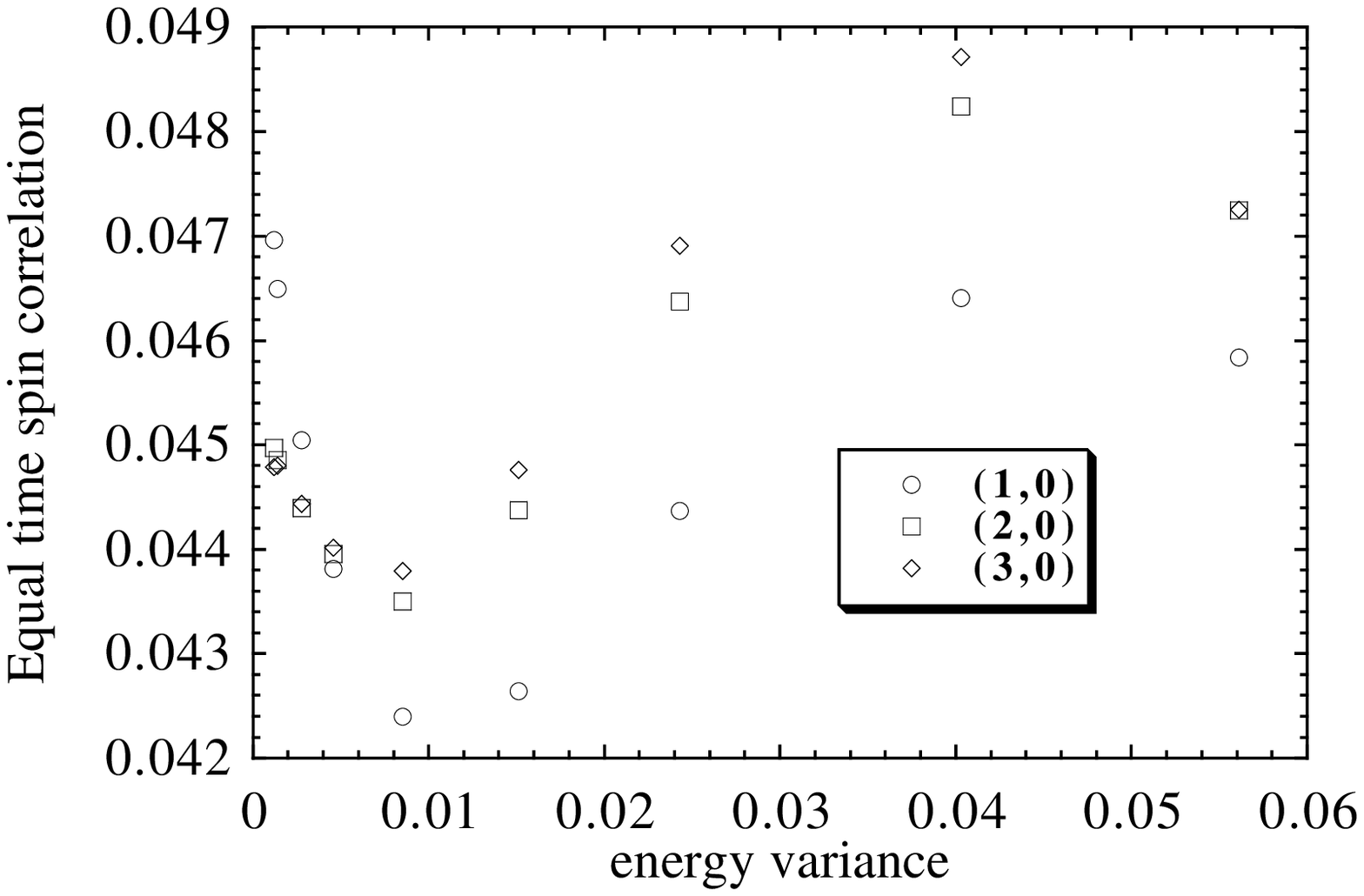}
 \caption{The extrapolation of the equal-time spin correlations
 $S\left(k_{x},k_{y}\right)$ to zero energy variance
 for the $6\times 2$ Hubbard model with 5 up 5 down electrons with the fully periodic boundary condition at $U/t=4.0$. 
 The wavenumber
 $(k_{x},k_{y})$ for each symbol is given in the inset. }
 \label{spin1}
\end{figure}

\begin{figure}
 \epsfxsize=80mm
 \epsffile{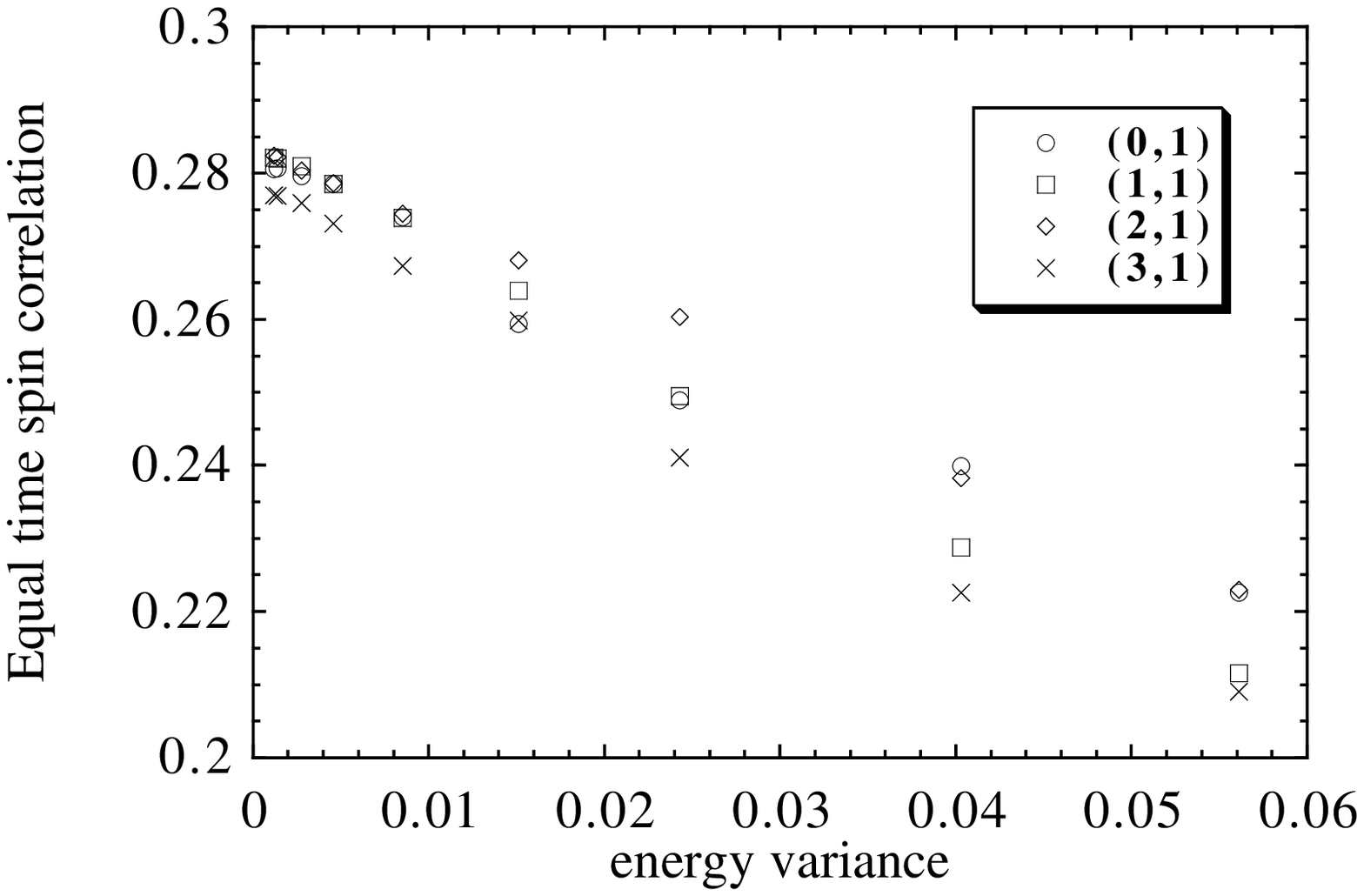}
 \caption{The extrapolation of the equal-time spin correlations
 $S\left(k_{x},k_{y}\right)$ to zero energy variance
 for the $6\times 2$ Hubbard model with 5 up 5 down electrons with the fully periodic boundary condition at $U/t=4.0$. The wavenumber
 $(k_{x},k_{y})$ for each symbol is given in the inset. }
 \label{spin2}
\end{figure}

Next, we evaluate the equal-time spin correlations and the momentum
distribution on $6\times 2$ Hubbard model, $5$ up $5$ down electrons with the fully periodic boundary condition at$U/t=4.0$. In our study, the equal-time spin correlations in
the momentum space is calculated from, 
\begin{equation}
 S\left(\mib{q}\right)=\frac{1}{3N}\sum_{i,j}^{N} 
  \left\langle\mib{S}_{i}\mib{S}_{j}
  \right\rangle e^{i\mib{q}\left(\mib{R}_{i}-\mib{R}_{j}\right)}
  \label{spinco}
\end{equation}
where $\mib{S}_{i}$ is the spin of the i-th site and each element of the
spin is calculated from 
\begin{eqnarray}
 S_{i}^{x}&=&\frac{1}{2}\left(S_{i}^{+}+S_{i}^{-}\right)
          =\frac{1}{2}\left(c_{i\uparrow}^{\dagger}c_{i\downarrow}
		             + c_{i\downarrow}^{\dagger}c_{i\uparrow}\right)
	  \nonumber\\
 S_{i}^{y}&=&\frac{1}{2i}\left(S_{i}^{+}-S_{i}^{-}\right)
          =\frac{1}{2i}\left(c_{i\uparrow}^{\dagger}c_{i\downarrow}
		             - c_{i\downarrow}^{\dagger}c_{i\uparrow}\right)
	  \nonumber\\
 S_{i}^{z}&=&\frac{1}{2}\left(n_{i\uparrow}-n_{i\downarrow}\right) .
  \label{detailspinco}
\end{eqnarray}
All the extrapolation behaviors of the equal-time spin correlations are
shown in Figs.\ref{spin1} and \ref{spin2}. The results after the
extrapolation procedure are shown in Table \ref{spin}. Here we take the
lattice constant to be unity. 

\begin{table}
 \caption{The equal-time spin correlations $S \left(k_{x},k_{y}\right)$
 for the $6\times 2$ Hubbard model with 5 up 5 down electrons with the fully periodic boundary condition at $U/t=4.0$}
 \begin{tabular}{@{\hspace{\tabcolsep}\extracolsep{\fill}}cccc}
  \hline
  $(k_{x},k_{y})$ & PIRG & exact diagonalization & relative error \\
  \hline
  (1,0) & 0.0482  & 0.0488 & 0.012 \\
  (2,0) & 0.0454  & 0.0458 & 0.0044 \\
  (3,0) & 0.0451  & 0.0457 & 0.013 \\
  (0,1) & 0.281   & 0.277  & 0.014 \\
  (1,1) & 0.283   & 0.281  & 0.0071 \\
  (2,1) & 0.284   & 0.286  & 0.0070 \\
  (3,1) & 0.278   & 0.279  & 0.0036 \\
  \hline
 \end{tabular}
 \label{spin}
\end{table}

\begin{figure}
 \epsfxsize=80mm
 \epsffile{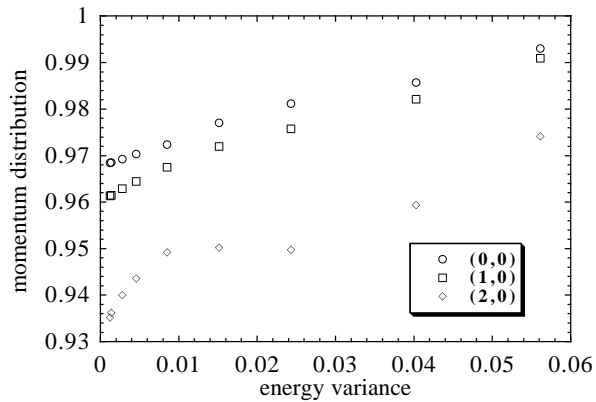}
 \caption{The extrapolation of the momentum distribution 
 $n\left(k_{x},k_{y}\right)$ to zero energy variance
 for the $6\times 2$ Hubbard model with 5 up 5 down electrons with the fully periodic boundary condition at $U/t=4.0$. The wavenumber
 $(k_{x},k_{y})$ for each symbol is given in the inset. }
 \label{mom1}
\end{figure}

\begin{figure}
 \epsfxsize=80mm
 \epsffile{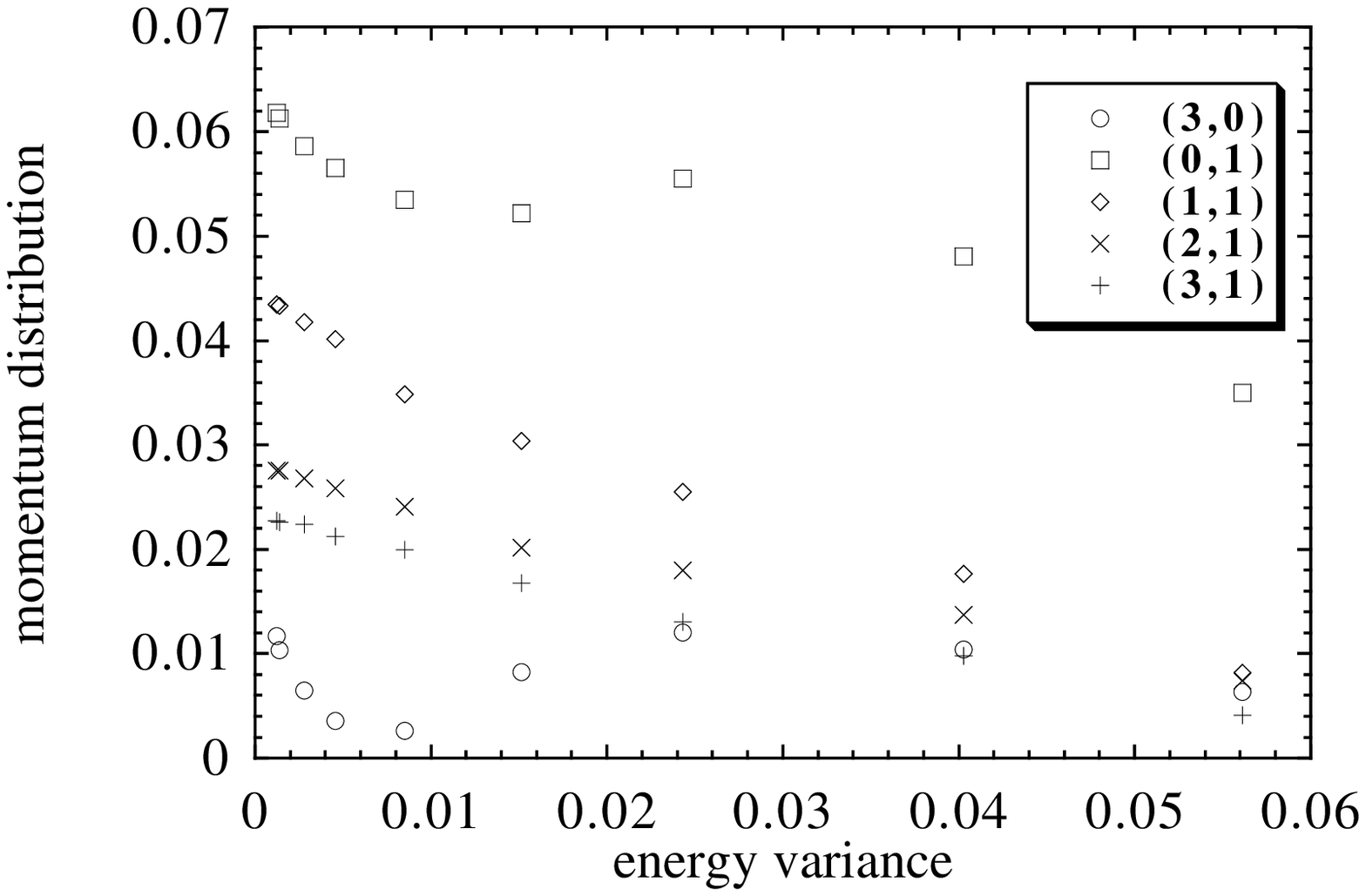}
 \caption{The extrapolation of the momentum distribution 
 $n\left(k_{x},k_{y}\right)$ to zero energy variance
 for the $6\times 2$ Hubbard model with 5 up 5 down electrons with the fully periodic boundary condition at $U/t=4.0$. The wavenumber
 $(k_{x},k_{y})$ for each symbol is given in the inset. }
 \label{mom2}
\end{figure}

The momentum distribution is calculated from,
\begin{equation}
 n \left(\mib{q}\right)=\frac{1}{2N}\sum_{i,j}^{N}\left\langle
    c_{j\uparrow}^{\dagger}c_{i\uparrow}+c_{j\downarrow}^{\dagger}c_{i\downarrow}
    \right\rangle e^{i\mib{q}\left(\mib{R}_{i}-\mib{R}_{j}\right)}
    \label{momdisEq}
\end{equation}
where $\mib{R}_{i}$ is the vector representing the place of the $i$-th
site. All the extrapolation behaviors of the momentum distribution are
shown in Figs.\ref{mom1} and \ref{mom2}. The comparison of the results
after the extrapolation is shown in Table \ref{mom}. 

Among the equal-time spin correlations and the momentum distribution, some
of them have relatively large errors of about a few percents.
Most of the relative errors are, however, less than 1 percent.

\begin{table}
 \caption{The momentum distribution $n \left(k_{x},k_{y}\right)$
 for the $6\times 2$ Hubbard model with 5 up 5 down electrons with the fully periodic boundary condition at $U/t=4.0$}
 \begin{tabular}{@{\hspace{\tabcolsep}\extracolsep{\fill}}cccc}
  \hline
  $(k_{x},k_{y})$ & PIRG & exact diagonalization & relative error \\
  \hline
  (0,0) & 0.9678  & 0.9681 & 0.00031 \\ 
  (1,0) & 0.9604  & 0.9601 & 0.00031 \\
  (2,0) & 0.9288  & 0.9281 & 0.00075 \\
  (3,0) & 0.02067 & 0.01850 & 0.092 \\
  (0,1) & 0.06515 & 0.06806 & 0.044 \\
  (1,1) & 0.04504 & 0.04513 & 0.0017 \\
  (2,1) & 0.02821 & 0.02787 & 0.012 \\
  (3,1) & 0.02294 & 0.02281 & 0.0057 \\
  \hline
 \end{tabular}
 \label{mom}
\end{table}

\section{Summary}
\label{conc}

Path-integral renormalization group(PIRG) is a numerical algorithm for
studying the ground state properties. 
The process filtering out the ground state $|\psi_{g}\rangle$ 
is performed in the imaginary time direction as,
\begin{displaymath}
 |\psi_{g}\rangle=\lim_{\tau \to \infty} \exp[-\tau \hat{H}]|\phi_{initial}\rangle .
\end{displaymath}
Therefore its formalism can be applied to any kind of systems and there
is no restriction on the spatial dimension of the system.
In this path-integral formalism, the ground state is represented by
chosen basis states $|\phi\rangle$,
\begin{displaymath}
 |\psi_{g}\rangle=\sum_{i}c_{i}|\phi_{i}\rangle .
\end{displaymath}
By the numerical renormalization, relevant basis states are selected and
irrelevant basis states are projected out.
This makes it possible to
calculate the approximate ground state directly as an optimized linear
combination of chosen basis states: 
\begin{displaymath}
 |\psi_{g}\rangle\approx\sum_{i=1}^{L}w_{i}|\phi_{i}\rangle .
\end{displaymath}
Therefore there is no negative sign problem even in frustrated systems.
In this way, PIRG can be applied to the systems which can not be treated
by existing algorithms such as the quantum Monte Carlo method or the density
matrix renormalization group. 

Because the converged state by PIRG is an approximate ground state under 
the restriction on the number of the basis states, the exact ground
state can 
be achieved by the extrapolation of the number of states $L$ to the
dimension of the whole Hilbert space. We have shown the general
extrapolation procedure in this paper:  
\begin{displaymath}
 \langle\hat{H}\rangle - \langle\hat{H}\rangle_{g} \propto \Delta E
\end{displaymath}
where $\Delta E$ is the energy variance,
\begin{displaymath}
 \Delta E=\frac{\langle\hat{H}^{2}\rangle - \langle\hat{H}\rangle^{2}}
  {\langle\hat{H}\rangle^{2}},
\end{displaymath}
$\langle\quad\rangle_{g}$, the expectation value in the true ground
state 
and $\langle\quad\rangle$, the expectation value in an approximate
ground state. This relation holds for sufficiently converged approximate
state. We confirm that the results of more than a hundred Slater
determinants follow the above relation and can be used for a linear
extrapolation in the Hubbard model. 
On the physical quantity $\hat{A}$, a similar
relation holds in most cases.
\begin{displaymath}
 \langle\hat{A}\rangle - \langle\hat{A}\rangle_{g} \propto \Delta E .
\end{displaymath}
For short-ranged correlation functions, the linearity holds at the same
level as the energy in the thermodynamic limit. 
By these extrapolation procedure, more accurate results are obtained 
while the variational principle is not strictly satisfied after the 
extrapolations. 
 
We compare the expectation values of the energy, the momentum
distribution and the equal-time spin correlations with those of
exact diagonalization on $6\times 2$ lattice systems and with those of
quantum Monte Carlo on larger systems. 
For the momentum distribution and the equal-time spin correlations,
the relative errors and differences from QMC are less than a few percents.   
Especially on the energy, the relative errors and differences 
from QMC have three digits accuracy. We confirm that these accuracy can
be achieved up to $12\times 12$ systems on the square lattice.  

We have also explained the dependence of the
computation time as $L^{2}N^{3}+L^{4}N$ where $N$ is the system size and
$L$, the dimension of the stored PIRG Hilbert subspace. We refer to some
implementation advice on PIRG and the efficiency of distributed memory
parallelization on PIRG methods. Because in general it is difficult to
parallelize efficiently the operations of matrices with the size such as
hundreds $\times$ hundreds or the iteration process, 
it is necessary to change the algorithm to 
make the convergence faster by projecting some basis states in parallel.       

In this study, we have dealt with Hubbard models using Slater
determinants as PIRG basis states. Applications of PIRG to other 
systems are very interesting and promising future projects.

\section*{Acknowledgments}
The authors thank F. F. Assaad for useful comments. 
This work is supported by the `Research for the Future` program from the Japan Society for the Promotion of Science under grant number
JSPS-RFTF97P01103. A part of our computation has been done at the
supercomputer center at the Institute for Solid State Physics,
University of Tokyo. 
\appendix
\section{Inner product of Slater determinants}
\label{DET}
Here we ignore the spin degrees of freedom for simplicity. 
A Slater determinant $|\phi_{a}\rangle$ is represented to be an 
$N\times M$ matrix $[\phi_{a}]$.
\begin{equation}
 |\phi_{a}\rangle=\prod_{j=1}^{M}
  \left(\sum_{i=1}^{N}[\phi_{a}]_{ij}c_{i\sigma}^{\dagger}\right)
  |0\rangle .
\end{equation}
where $a$ is a symbol to distinguish Slater determinants. 
Eq.(\ref{determinant}) can be obtained as the following:
\begin{eqnarray}
 \langle\phi_{a}|\phi_{b}\rangle &=& \langle 0|\prod_{j=1}^{M}\prod_{l=1}^{M}
  \left(\sum_{i=1}^{N}\left[\phi_{a}\right]_{ij}c_{i}\right)
  \left(\sum_{k=1}^{N}\left[\phi_{b}\right]_{kl}c_{k}^{\dagger}\right)
  |0\rangle\\
 &=&
 \sum_{\mib{S}}^{{}_{N}C_{M}}
 \left\{\sum_{\mib{n}}^{M!}\sum_{\mib{m}}^{M!}sgn(\mib{n})sgn(\mib{m})
		\prod_{j=1}^{M} 
		\left(\left[\phi_{a}\right]_{\mib{S}_{\mib{n}(j)}j}
		 \left[\phi_{b}\right]_{\mib{S}_{\mib{m}(j)}j} \right)\right\} .
    \label{B3}
\end{eqnarray}
Here $\mib{n}$ and $\mib{m}$ are permutation of $M$ symbols:
\begin{eqnarray}
 \mib{n}&=&
  \left(
   \begin{array}{lll}
 1,\qquad 2,\cdots ,\qquad M\\
 \mib{n}(1),\mib{n}(2),\cdots ,\mib{n}(M)
   \end{array}
 \right) ,
\end{eqnarray}
the same is assumed for $\mib{m}$; $\sum_{\mib{n}}^{M!}$ means taking
the sum over all permutations of 
$M$ symbols; $sgn(\mib{n})$ is the signature of the permutation $\mib{n}$:
\begin{eqnarray}
  sgn(\mib{n})=
   \left\{
    \begin{array}{lll}
     +1\qquad\textrm{for an even permutation}\\
     -1\qquad\textrm{for an odd permuation}
    \end{array}
    \right. ;
\end{eqnarray}
$\mib{S}$ is a set of $M$ numbers chosen from $N$ numbers $1,2,\cdots
,N$ and we assume the ascendent order on this set $\mib{S}$; 
$\sum_{\mib{S}}^{{}_{N}C_{M}}$ means taking the sum over all
combinations of $M$ numbers. Then the inner product can be transformed
from Eq.(\ref{B3}) as the following:
\begin{eqnarray}
 \langle\phi_{a}|\phi_{b}\rangle 
  &=& M!\sum_{\mib{S}}^{{}_{N}C_{M}}\left\{\sum_{\mib{n}}^{M!}sgn(\mib{n})
  \prod_{j=1}^{M}\left(\left[\phi_{a}\right]_{\mib{S}_{j}j}
		\left[\phi_{b}\right]_{\mib{S}_{\mib{n}(j)}j}\right)\right\}
   \\
  &=& M!\sum_{\mib{S}}^{{}_{N}C_{M}}\left\{\sum_{\mib{n}}^{M!}sgn(\mib{n})
  \prod_{j=1}^{M}\left(\left[\phi_{a}\right]_{\mib{S}_{j}j}
     \left[\phi_{b}\right]_{\mib{S}_{j}\mib{n}(j)}\right)\right\}\\
 &=& \sum_{\mib{n}}^{M!}sgn(\mib{n})\prod_{j=1}^{M}
  \left(\sum_{i=1}^{N}\left[\phi_{a}\right]_{ij}
 \left[\phi_{b}\right]_{i\mib{n}(j)}\right)\\
 &=& \textrm{det}\left({}^{t}\left[\phi_{a}\right]\left[\phi_{b}\right]\right) .
\end{eqnarray}



\begin{thebibliography}{999}
%
%
\bibitem{QMC1}M.Imada and Y.Hatsugai:
J.Phys.Soc.Jpn.{\bf 58}(1989)3752.
%
\bibitem{QMC2}N.Furukawa and M.Imada:
J.Phys.Soc.Jpn.{\bf 61}(1992)3331.
%
\bibitem{DMRG}S.R.White:
Phys.Rev.{\bf B48}(1993)10345.
%
\bibitem{TDM1}T.Nishino:
J.Phys.Soc.Jpn.{\bf 64}(1995)3598.
%
\bibitem{TDM2}N.Shibata:
J.Phys.Soc.Jpn.{\bf 66}(1997)2221.
%
\bibitem{TDM3}N.Shibata:
Phys.Rev.{\bf B56}(1997)5061.
%
\bibitem{PIRG}M.Imada and T.Kashima:
J.Phys.Soc.Jpn.{\bf 69}(2000)2723.
%
\bibitem{Sorella}S.Sorella:
preprint, cond-mat/0009149
%
\bibitem{TrEx1}P.Prelov\v{s}ek and X.Zotos:
Phys.Rev.{\bf B47}(1993)5984
%
\bibitem{TrEx2}J.Riera and E.Dagotto:
Phys.Rev.{\bf B48}(1993)9515
%
\bibitem{TrEx3}N.A.Modine and E.Kaxiras:
Phys.Rev.{\bf B53}(1996)2546
%
\bibitem{MC_Tr1}H.D.Raedt adn W.v.d.Linden:
Phys.Rev.{\bf B45}(1992)8787
%
\bibitem{MC_Tr2}M.Honma, T.Misusaki and T.Otsuka:
Phys.Rev.Lett{\bf 75}(1995)1284
\end{thebibliography}
\end{document}